# THE PHOTOPROCESSES IN THE Nd AND $d^3He$ SYSTEMS ON THE BASIS OF POTENTIAL CLUSTER MODELS

S. B. Dubovichenko

Kazakh State National University, Almaty, Kazakstan

Direct cluster disintegration and radiative capture in the pd, nd, and $d^3He$ channels are considered on the basis of potential two-cluster models. Interaction between clusters is described by potentials that, in some cases, involve forbidden states and which are fitted to the phase shifts of elastic scattering at low energies. Phase shifts and potentials of scattering are separated according to the Young orbital diagrams.

Previously, we calculated the total cross sections for photodisintegration and radiative capture in the $^4He^3H$, $^4He^3He$, $d^4He$, and $^3He^3H$ systems on the basis of two-cluster models [1] with potentials involving forbidden states (FS) [2]. Such intercluster interactions describe the experimental phase shifts of elastic scattering at low energies, whileFSs make it possible to take into account the Pauli exclusion principle effectively in interactions between clusters [3] without using a repulsive core. It was shown in [2] that certain features of the $^6Li$ and $^7Li$ nuclei, in which the probability of clustering in the above channels is considered to be relatively high, can be reproduced on the basis of cluster models involving FSs. The states in such systems possess pure orbital symmetry [3], and the potentials derived from experimental data on the phase shifts of elastic scattering can be used directly to describe the characteristics of the ground states (GS) of nuclei [2].

In systems formed by lighter clusters such as Nd, $N^3H$, $d^3He$, etc., mixing in orbital symmetries is observed for states with minimum values of spin; hence, it is necessary to extract pure components applicable to the analysis of GS characteristics from interactions obtained on the basis of experimental phase shifts of elastic scattering [4]. For example, the spin of the Nd system can assume values of 1/2 and 3/2, while the isospin is 1/2. The spin and isospin wave functions (WF) are characterized by definite Young diagrams $\{f\}_S$ and $\{f\}_T$, which prescribe their symmetry under permutations of nucleon coordinates. For example, the spin symmetry is characterized by the diagrams $\{3\}_S$ for $S = 3/2$ and $\{21\}_S$ for $S = 1/2$, while the isospin symmetry is specified by the diagram $\{21\}_T$ As the symmetry of the spin-isospin wave function is determined by the direct inner product $\{f\}_{ST} = \{f\}_S \times \{f\}_T$ [4], we have $\{f\}_{ST} = \{1^3\}+\{21\} + \{3\}$ in the doublet channel. In the quartet channel, this product yields only one diagram $\{21\}_{ST}$. The symmetry of the total WF, including its orbital component, is determined in a similar way: $\{f\}=\{f\}_L \times \{f\}_{ST}$. The total WF of the system does not vanish identically as the result of antisymmetrization only if it contains the antisymmetric component $\{1^N\}$, which is realized by multiplying the conjugate diagrams $\{f\}_L$ and $\{f\}_{ST}$. For this reason, the diagrams $\{f\}_L$ conjugate to $\{f\}_{ST}$ are considered to be allowed in the given channel. Using the Littlewood theorem, we can determine the possible Young orbital diagrams in a system of $N = n_1 + n_2$ particles as the direct outer product of the orbital diagram in each subsystem. This yields $\{f\}_L = \{2\}\}\{1\}$

= $\{21\}_L + \{3\}_L$ in the case under study. The diagram $\{2\}$ corresponds to a deuteron with LS = 01. It follows that the only allowed orbital WF in the quartet channel has the symmetry $\{21\}_L$ while the function $\{3\}_L$ is forbidden because the product $\{21\}_{ST} \times \{3\}_L$ does not lead to an antisymmetric WF component. In the doublet channel, we have the products $\{3\}_L \times \{1^3\}_{ST} = \{1^3\}$ and $\{21\}_{ST} \times \{21\}_L \sim \{1^3\}$, which lead to antisymmetric diagrams. Thus, the two possible Young orbital diagrams $\{21\}_L$ and $\{3\}_L$ are allowed in the doublet channel.

This result leads to the concept of mixing in orbital diagrams for states with definite values ST=1/2 1/2 for any L. Consequently, the S-wave doublet interaction potential obtained from experimental phase shifts of scattering depends effectively on both orbital diagrams, while the ground state (for example, of the triton) corresponds to the pure symmetry $\{3\}_L$. This means that these potentials are different in an Nd system, and we must extract from the potential $V^{\{3\}+\{21\}}$ describing scattering the component $V^{\{3\}}$, which can be used, in principle, to calculate the GS characteristics [4]. In particular, the analysis of photonuclear processes must be based on potentials corresponding to pure Young diagrams for bound states (BS) and on potentials reconstructed directly from experimental phase shifts for scattering states, for example, final states of photodisintegration.

In order to construct an interaction corresponding to pure Young diagrams, we must extract pure phases with the diagram $\{3\}_L$ from experimental phase shifts of scattering. According to [4, 5], experimental phase shifts of the doublet channel can be represented as the half-sum of pure phases with the diagrams $\{21\}_L$ and $\{3\}_L$. If we assume that quartet phases corresponding to the pure orbital diagrams $\{21\}_L$ can be used as doublet phases with $\{21\}_L$, we can easily find doublet phases with the pure diagrams $\{3\}_L$ and parametrize pure interaction with the aid of these phases. Such Nd interactions were derived in [4,5], where it was shown that the binding energies in cluster channels and the asymptotic constant can be reproduced correctly. Reasonable estimates for the charge radius and the elastic Coulomb form factor at small momentum transfers are also obtained if small deformations of the deuteron cluster are taken into consideration. The calculations in [4] were based on the potential cluster model in which the nucleus is represented as a system of two structureless fragments whose properties are assumed to be close to those of the corresponding free particles. Explicit antisymmetrization of the wave function of the system has not been carried out, but purely attractive interactions between clusters sometimes involve FSs. In the presence of FSs, the resulting WF of the relative motion of clusters oscillates in the interior of the nucleus. The corresponding scattering phase shifts obey the generalized Levinson theorem and tend to zero from above at high energies [3,4].

The mixing of orbital Young diagrams is typical not only of the Nd system. States with minimum values of spin can have mixed orbital symmetries not only in all light cluster systems of the type Nd, $p^3H$, $n^3He$, dd, and $d^3He$ ($d^3H$) [5], but also in heavier systems such as $N^6Li$, $N^7Li$, and $d^6Li$ [4]. In all cases, the experimental mixed phase shift of scattering can be represented as the half-sum of pure phases. Detailed classification of FSs and allowed states (AS) for such cluster systems and heavier systems is given in [4].

The analysis of differential cross sections for photo-nuclear processes in the

Nd and d³He systems for potentials with FSs and separation in orbital diagrams was successfully carried out in [6]. However, the total cross sections for such systems in cluster models with FSs and with the separation of Young diagrams have not been considered.

Cross sections for radiative capture in the long-wave approximation are calculated here on the basis of the well-known expression [1,7]. The Gaussian potential used to describe interactions between clusters has the form [4]

$$V(r) = -V_0 \exp(-\alpha r^2) + Z_1 Z_2/r$$

The parameters of the potentials were chosen in such a way that one set of parameters described phase shifts of partial waves with the same parity in the low-energy region. It was found in [4] that $V_S^{\{3\}} = 34.8$ MeV, $\alpha = 0.15$ fm$^{-2}$ $V^{\{3\}+\{21\}} = -0.4$ MeV, and $\alpha = 0.01$ fm$^{-2}$ (purely repulsive potential). For even scattering phase shifts, we have $V_0 = 35$ MeV and $\alpha = 0.1$ fm$^{-2}$. The mass values for the various clusters were taken to be $M_d = 2.0135537$, $M_p = 1.0072766$, and $M_n = 1.0086653$.

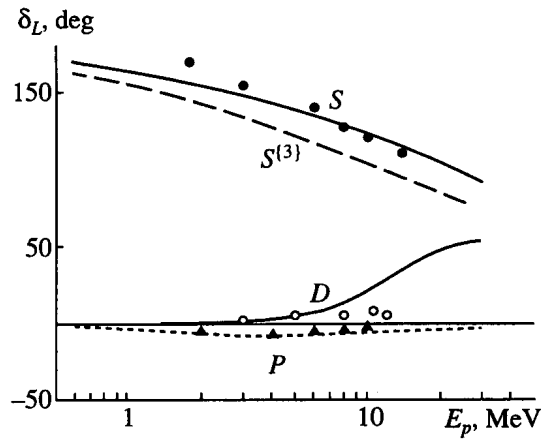

Fig. 1. Doublet and pure phase shifts of elastic pd scattering. The experimental data are taken from [8].

For such values, a pure S-wave potential leads to binding energies of -5.51 MeV in the pd channel and -6.26 MeV in the nd channel (cf. the experimental values -5.493 and -6.257 MeV, respectively). The asymptotic constant in both systems was found to be 2.30 (5). This value is in accord with known results [8] lying in the interval 1.6 - 2.05 for the nd system and in the interval 1.9 - 2.3 for the pd system. Figure I shows calculated phases for S-, P-, and .D-wave interactions at energies up to 30 MeV. The experimental data are taken from [9]. The allowed BS in the mixed S-wave interaction corresponds to an energy of -9.2 MeV, which is higher than the binding energy of the ³He nucleus. The parameters of this interaction differ from those presented in [4]. The potential used here describes the D-wave phase shifts more correctly. This circumstance is important for the analysis of E2 photo-nuclear processes.

Total cross sections were calculated for El and E2 transitions due to the orbital component of the electric operator Q(L). The magnetic cross sections and the cross sections determined by the spin component of the electric operator were found to be relatively small. E1 transitions in the Nd system are allowed between the pure ground $^2S$ state and the $^2P$ scattering state.

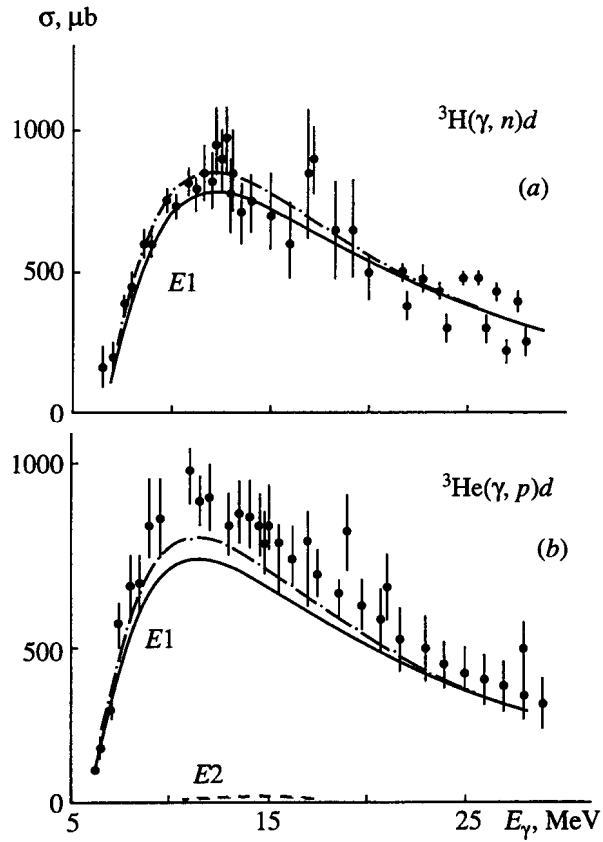

Fig. 2. Total cross sections of photodisintegration in the processes $^3H(\gamma,n)d$ (a) and $^3He(\gamma,p)d$ (b). The solid and dot-and-dash curves correspond to E1 cross sections calculated with different GS potentials. The experimental data are taken from [10,11].

The cross sections for photonuclear processes are proportional to the factor $(Z_1/M_1 +(-1)^J Z_2/M_2)$, which has the same magnitude for the nd and pd systems in E1 transitions and differs by almost an order of magnitude from the corresponding quantities for El processes. Therefore, the cross section for the El process is noticeable only for the pd system; this, however, does not explain the difference in the experimental values of the cross sections for pd and nd photodisintegration.

Figure 2 shows calculated total cross sections for the photodisintegration of $^3H$ and $^3He$ nuclei through the nd and pd channels. The experimental data are taken from [10, 11]. It can be seen from Fig. la that the cross section for disintegration through the nd channel is explained completely by the E1 transition. At the same time (see Fig. lb), the E1 cross section for pd disintegration (solid curve) lies considerably below the experimental results, and even the inclusion of El processes (dashed curve) does not provide a correct description of the experimental cross section. Figure 3 shows radiative-capture cross sections for the pd system and the experimental results from [12]. The solid curve corresponds to the cross section for the E1 process

The astrophysical S factor associated with the E1 transition is shown in Fig. 3b. The experimental results were obtained by recalculating the data from [12]. Linear extrapolation of the S(E1) factor to zero energy gives a value of $1.3(2) \times 10^{-4}$ keV b. The cross sections for the processes under consideration are virtually independent of the magnitude of the P-wave scattering potential. Even if this potential is set equal to zero, the shape of the cross section does not change, while its magnitude varies by a few percent. However, the cross sections depend strongly on the magnitude and shape of the GS interaction. For example, if we use the potential with the parameters 31.93 MeV and 0.13 fm$^{-2}$ (these values yield virtually the same BS characteristics and phases), the cross sections for photonuclear processes become considerably larger, and the theoretical values of the pd cross sections fall within the region of experimental errors (shown in Figs. 2 and 3 by dot-and-dash curves).

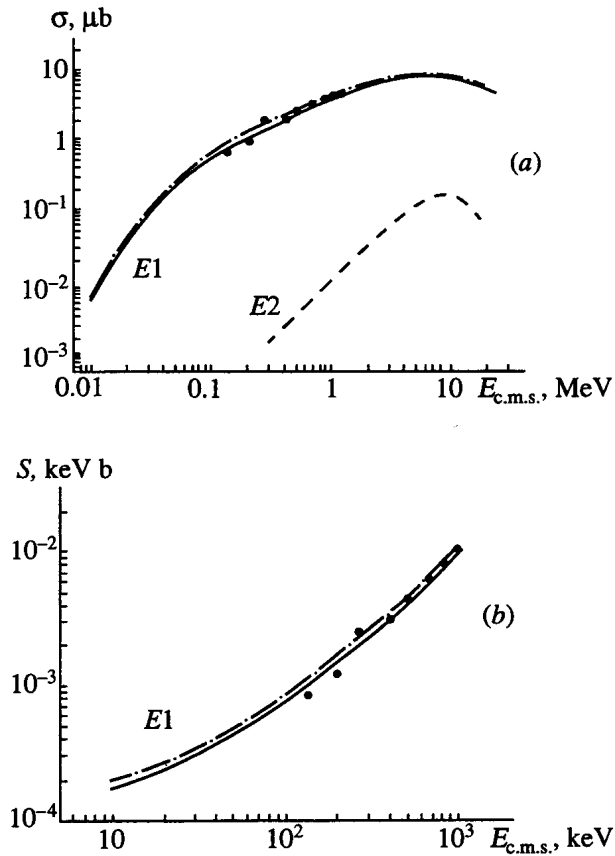

Fig. 3. (a) Total cross sections for the radiative capture d(p, $\gamma^3$)He and (b) the astrophysical S factor. The solid and dot-and-dash curves correspond to E1 cross sections calculated with different GS potentials. The experimental data are taken from [12].

The parameters of the potential cannot be determined unambiguously from pure phases. For this reason, the GS characteristics were used in [4] to accomplish this end. However, this procedure also yields several versions of interaction that correctly reproduce the binding energy and the asymptotic constant and which make it possible to determine the charge radius and the elastic Coulomb form factor for small momentum transfers, provided that possible deformations of the deuteron cluster are taken into account. Among other things, two sets of such parameters are presented in [4]. Hence, the form of the potential can be determined more precisely from the cross sections for photonuclear processes in cases where these cross sections are highly sensitive to the GS WF. It was shown above that a change of about 10% in the parameters from the values presented in [4] provides quite a reasonable description of the experimental cross sections for photodisintegration at moderate energies. Such a potential leads to a binding energy of 5.5 MeV, an asymptotic constant of 2.45(10), and a charge radius of 1.83 fm for 40% deformation of the deuteron. Hence, it can be assumed that the model used for potentials matched with the scattering phase shifts in the continuous spectrum and with GS characteristics with allowance for deuteron deformation can be used to describe the experimental data, including S-factors in the astrophysical region, within acceptable error limits.

The cross sections for these processes have been calculated in a number of studies on the basis of different models. In particular, the total cross sections at moderate energies can be described quite correctly in some versions of the method of hyperspherical functions [11]. However, such approaches usually do not take into account the supermultiplet symmetry of WFs with the separation of Young diagrams, which makes it possible to analyze the structure of interaction between clusters and to determine the existence and position of allowed and forbidden states. Such an analysis yields not only potentials that correctly describe the energy dependence of the scattering phase shifts, but also the WFs of BSs and the continuum that have a definite structure of nodes in the interior of the nucleus. In contrast to core potentials, such WFs do not die out at small distances, and the positions of their nodes differ from those characteristic of functions of purely attractive potentials without FSs.

Let us now use Young diagrams to construct the classification of states of the five-nucleon system $d^3$He or $d^3$H, which may exist in two spin states with S = 1/2 and 3/2 with the diagrams $\{32\}_S$ and $\{41\}_S$ respectively, for an isospin of 1/2 with the diagram: $\{32\}_S$. The spin-isospin WF is characterized by the following diagrams: $\{32\} \times (32) = \{5\} + \{41\} + \{32\} + \{311\} + \{221\} + \{2111\}$ for S = 1/2 and $\{32\}_T \times \{41\}_S = \{41\} + \{32\} + \{311\} + \{221\}$ for S = 3/2 [4]. The possible orbital diagrams determined from the Littlewood theorem are $\{3\} \times \{2\} = \{5\}_L + \{4\}_L + \{32\}_L$. It immediately follows that two possible orbital diagrams $\{41\}_L$ and $\{32\}_L$ are allowed for S = 1/2 and any L because the spin-isospin component of the WF contains the conjugate symmetries $\{2111\}$ and $\{221\}$. Because of the absence of the symmetry $\{2111\}$ conjugate to $\{41\}_L$ from the spin-isospin WF, only the $\{32\}_L$ symmetry is allowed for S = 3/2 and arbitrary L. Components of the radial WF with $\{5\}_L$ are forbidden in all spin channels.

Hence, the $^2$P wave for a channel with minimum spin contains an allowed BS with the diagram $\{41\}_L$ corresponding to the ground states of the nuclei $^5$He-$^5$Li. The S wave involves a forbidden BS with $\{5\}_L$. The bound FS in the quartet P wave corresponds to the symmetry $\{41\}_L$.

Experimental data on the phase shifts of $d^3$He scattering are available only for the S wave in the energy region up to 3 MeV [13]. In constructing the potential, we therefore supplemented these data with the results of multichannel analysis carried out with aid of the resonating-group method (RGM) at energies up to 20 MeV [14].

The S-wave component of interaction was fixed by fitting the experimental phase shifts at energies up to 3 MeV. These potentials were used to calculate the phase shifts for L = 0 at higher energies and the D-wave phase shifts over the entire energy range. Because experimental data on P-wave phase shifts are not available, this part of the potential was constructed entirely on the basis of RGM data. According to the results of analysis carried out in [4], the parameters of interaction for even doublet phase shifts are as follows: $V_0$ = 45.5 MeV and $\alpha$ = 0.15 fm$^{-2}$. For odd doublet phase shifts, we have $V_0$ = 61 MeV and the same value of $\alpha$. Figure 4 shows the calculated doublet S-, P-, and D-wave phase shifts, along with the results obtained in [13, 14]. It can be seen that the energy dependence of P-wave phase shifts is described correctly over the entire energy range. The phase shifts with L = 2 do not agree with RGM calculations [14], but reproduce the experimental data more correctly because the D resonance is located at an energy of 3.6 ± 0.5 MeV relative to the cluster-channel threshold and has a width of about 5 MeV [15]. The forbidden state with LS = 0 1/2 was found to be at an energy of -15.9 MeV, while the energy corresponding to the forbidden state with LS = 03/2 is -11.8 MeV. In the P-wave with S = 3/2, the potential leads to an FS at an energy of -1.0 MeV.

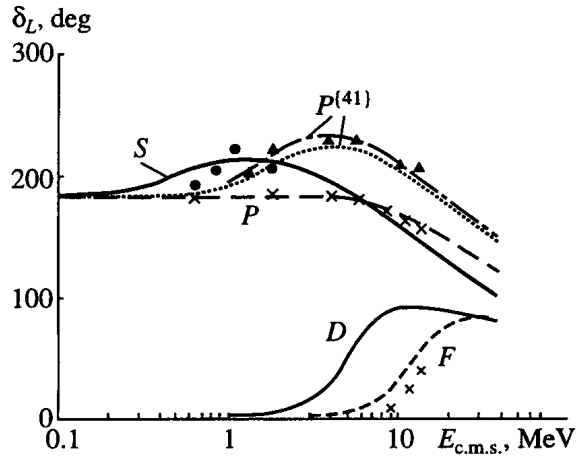

Fig. 4. Doublet and pure phase shifts of elastic d$^3$He scattering: solid curves - S- and D-wave phase shifts, short dashes - F-wave phase shifts, long dashes - P -wave phase shifts, dotted and dot-and-dash curves - pure P-wave phase shifts for potentials of depth 75.5 and 78.0 MeV, experimental data from [13] for the S-wave phase shifts, (x) RGM calculations for P- and D-wave phase shifts, and (Δ) pure phase shifts obtained on the basis of RGM calculations for the doublet and quartet phases.

Separating experimental phase shifts of scattering in accordance with the Young diagrams as in the case of the Nd system, we can obtain pure phases with {41}$_L$ and thereby determine the pure doublet potential, whose parameters are as follows: $V_0$ = 42 MeV for even waves and 78 MeV for odd waves and $\alpha$ = 0.15 fm$^{-2}$ in both cases. The results of calculations for the pure P-wave phase shift are presented in Fig. 4 by the dot-and-dash curve. The triangles correspond to the pure phase shifts obtained from the RGM phase shifts of scattering. It can be seen that the theoretical pure P-wave phase shift successfully describes these data. The al-lowed BS in the P wave lies at -17.7 MeV, which is in reasonable agreement with the data obtained in [15], where the $P_{3/2}$ level is found to occur at an energy of -16.4 MeV and has a width of about 1.5 MeV. In order to obtain the correct value of the 3/2 - level energy, the depth of interaction in the P wave must be changed insignificantly. This is equivalent to the inclusion of spin-orbit splitting. We must assume that $V_0 = 75.5$ MeV, with the same geometry of the potential. This potential is used in subsequent calculations, as the cross sections are very sensitive to the energy of emitted photons even for E1 transitions. This change in the potential depth leads to an insignificant variation of the phase shift (dotted curve in Fig. 4).

The $P_{1/2}$-level energy amounts to 9 MeV, its width being on the order of 5 MeV [15]. For capture from the D wave, the value of $^2P_J$ is twice as large as that in the case of S capture (it should be borne in mind that spin-orbit splitting does not occur in this potential). The BS energy obtained from the pure potential coincides with the level energy, and the corresponding cross section is calculated to a higher degree of accuracy than the cross section for capture to the $P_{1/2}$ level because the same potential is used in both cases. However, in view of the difference in the energies of the levels, the cross section for capture to the 1/2⁻ level is an order of magnitude smaller than the cross section for capture to the 3/2⁻ level.

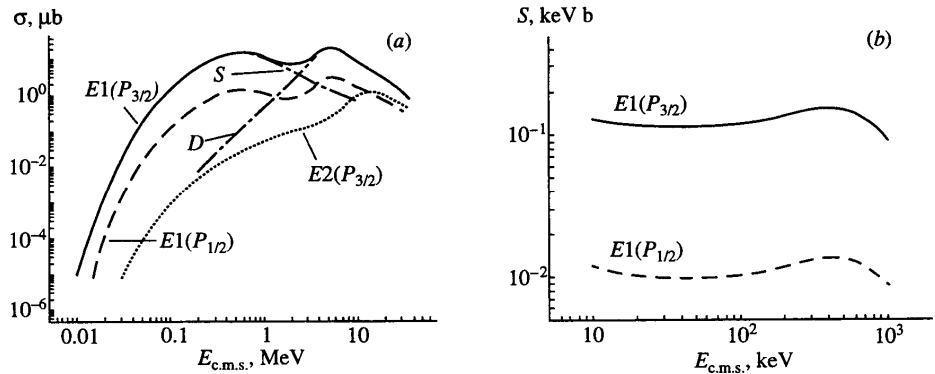

Fig. 5. (a) Total cross sections for the radiative capture d($^3$He, γ)$^5$Li: solid curve - cross section for the E1 transition to the $P_{3/2}$ level, dotted curve - cross section for the E2 transition to the $P_{3/2}$ level, dashed curve - cross section for the E1 transition to the level, and dot-and-dash curves - contributions from the S and D waves to the cross section for the E1 transition to the $P_{3/2}$ level, (b) The astrophysical S factor: the solid and dashed curves represent the S factors for the E1 transitions to the levels $P_{3/2}$ and $P_{1/2}$, respectively.

Figure 5a shows the calculated cross section for radiative capture for the d$^3$He system in the energy range from 10 keV to 40 MeV. The two peaks observed in the cross section for capture to both P levels at energies 0.5-0.7 and 5 - 6 MeV are due to resonances in the $^2$S and $^2$D waves, respectively, and correspond approximately to experimental levels with energies of 1.6 ± 1 MeV at $J = 1/2^+$ and 3.6 ± 0.5 MeV at $J = 3/2^+(5/2^+)$ [15]. These cross sections are shown in the figure by dot-and-dash curves. The dotted curve represents the calculated cross section for the E2 process from scattering P and F waves to the $P_{3/2}$ level. The analogous cross section for the $P_{1/2}$ state is an order of magnitude smaller. The energies of theoretical and experi-mental resonances in the S and D waves are in comparatively good agreement; for example, the width of the D level is about 5 MeV [15]. Figure 5b shows the astrophysical S factor for El transitions to the $P_{3/2}$ and $P_{1/2}$ levels. Linear extrapolation of the S factors to zero energy yields $S(E1, 3/2) = 1.5(3) \times 10^{-1}$ keV b and $S(E1, 1/2) = 1.5(3) \times 10^{-2}$ keV b.

Thus, the application of the one-channel potential cluster model involving forbidden states and splitting in accordance with Young diagrams to the analysis of light nuclei makes it possible to describe the available experimental data on the total cross sections for photodisintegration and radiative capture in the Nd system at moderate energies and to obtain quantitative results for the $d^3He$ system. The astrophysical S factor in the $d^3He$ system proves to be three orders of magnitude higher than that in the pd system because of the presence of an over-the-threshold resonance in the S wave. The parameters of the potentials employed here were preset using the phase shifts of elastic scattering of clusters and the GS characteristics. In this sense, we can speak of a unified description of the BS characteristics and the cross sections for photonuclear processes.